\documentclass[11pt]{article}

\usepackage[a4paper,margin=28mm]{geometry}
\usepackage[T1]{fontenc}
\usepackage{lmodern}
\usepackage{microtype}
\usepackage{amsmath,amssymb,amsthm,mathtools}
\usepackage{bm}
\usepackage{mathrsfs}
\usepackage{aliascnt}
\usepackage{xcolor}
\definecolor{revisionpurple}{RGB}{123,42,174}
\newcommand{\rev}[1]{{\color{black}#1}}
\usepackage{hyperref}
\usepackage[nameinlink,noabbrev]{cleveref}

\hypersetup{
  colorlinks=true,
  linkcolor=blue!55!black,
  citecolor=blue!55!black,
  urlcolor=blue!55!black,
  pdftitle={Absence of Excited Bound States for All Fibers of the Froehlich Polaron at Weak Coupling},
  pdfauthor={David Mitrouskas}
}

\allowdisplaybreaks
\numberwithin{equation}{section}

\newtheorem{theorem}{Theorem}[section]
\newaliascnt{proposition}{theorem}
\newtheorem{proposition}[proposition]{Proposition}
\aliascntresetthe{proposition}
\newaliascnt{lemma}{theorem}
\newtheorem{lemma}[lemma]{Lemma}
\aliascntresetthe{lemma}
\newaliascnt{corollary}{theorem}
\newtheorem{corollary}[corollary]{Corollary}
\aliascntresetthe{corollary}
\theoremstyle{definition}
\newaliascnt{remark}{theorem}
\newtheorem{remark}[remark]{Remark}
\aliascntresetthe{remark}

\crefname{theorem}{theorem}{theorems}
\Crefname{theorem}{Theorem}{Theorems}
\crefname{proposition}{proposition}{propositions}
\Crefname{proposition}{Proposition}{Propositions}
\crefname{lemma}{lemma}{lemmas}
\Crefname{lemma}{Lemma}{Lemmas}
\crefname{corollary}{corollary}{corollaries}
\Crefname{corollary}{Corollary}{Corollaries}
\crefname{remark}{remark}{remarks}
\Crefname{remark}{Remark}{Remarks}

\newcommand{\F}{\mathcal F}
\newcommand{\R}{\mathbb R}
\newcommand{\one}{\mathbf 1}
\newcommand{\Pf}{P_{\mathrm f}}
\newcommand{\Phiop}{\Phi}
\newcommand{\eps}{\varepsilon}
\newcommand{\norm}[1]{\lVert #1\rVert}
\newcommand{\dd}{\,\mathrm d}

\title{\huge{Absence of Excited Bound States for All Fibers of\\ the Fröhlich Polaron at Weak Coupling}}
\author{David Mitrouskas\thanks{{david.mitrouskas@gmail.com}}}
\date{}

\begin{document}
\begingroup
\makeatletter
\renewcommand{\@fnsymbol}[1]{}
\makeatother
\maketitle
\endgroup
\frenchspacing 
\begin{abstract}
We study the discrete spectrum of the three-dimensional
Fr\"ohlich polaron at weak coupling and fixed total momentum.
We prove that, for sufficiently small coupling and every
total momentum $0<|\xi|\le1/4$, the ground-state
eigenvalue is the only eigenvalue below the essential
spectrum. Our proof builds on the analogous zero-momentum
result of \cite{Seiringer2023}. Combined with that result
and the work of \cite{Dahlbaek2017}, this implies that,
at sufficiently weak coupling, no fiber Hamiltonian
has a discrete excited eigenvalue.
\end{abstract}

\section{Main result}\label{sec:introduction}

The Fr\"ohlich model describes a particle interacting with a quantized field of
optical phonons.  
After decomposition with respect to the conserved total momentum
$\xi\in\R^3$, the Fröhlich Hamiltonian is
represented by the family of fiber Hamiltonians
\begin{align}
H_g(\xi)=(\Pf-\xi)^2+N+\Phiop(v_g),
\qquad
v_g(k)=\frac{g}{|k|},
\label{eq:model}
\end{align}
acting on the bosonic Fock space $\F=\F_{\mathrm s}(L^2(\R^3))$.
Here $g\in\R$ denotes the coupling constant, while
$\Pf=\int dk\, k\,a_k^*a_k$ and $N=\int dk\,a_k^*a_k$
are the field momentum and number operators, respectively.
The interaction is given by the field operator $\Phiop(v)=a^*(v)+a(v)$,
where $a^*(v)=\int dk\,v(k)a_k^*$, and the creation and
annihilation operators satisfy the canonical commutation
relations $[a_k,a_\ell^*]=\delta(k-\ell)$.

Since $v_g$ is not square-integrable, the interaction in \eqref{eq:model} is understood in the quadratic-form sense. The Lieb--Yamazaki estimate \eqref{eq:form-bound-full} implies that the quadratic form associated with \eqref{eq:model} is closed and bounded from below and therefore defines a self-adjoint operator. See \cite{LiebYamazaki1958,GriesemerWuensch2016} for details. The goal of this article is to study the discrete spectrum of this operator.

Conjugation by $(-1)^{N}$ maps $H_g(\xi)$ to $H_{-g}(\xi)$, and hence
the spectrum depends only on $|g|$. We write
\begin{equation}\label{eq:energies}
 E_g(\xi)=\inf\sigma(H_g(\xi)),
 \qquad
 E_0=E_g(0)
\end{equation}
for the ground state energy at fixed momentum. It is known \cite{LampartMitrouskasMysliwy2023,
Polzer2023}  that $E_0<E_g(\xi)$ for every $\xi\neq0$. Moreover, the essential spectrum is given by
\begin{equation}\label{eq:threshold}
\sigma_{\mathrm{ess}}(H_g(\xi))=[E_0+1,\infty) \qquad \text{for all}
\qquad \xi\in\R^3.
\end{equation}
We refer to \cite{Moller2006} and \cite[Lemma~1.1]{LampartMitrouskasMysliwy2023}. Both statements hold for all values of $g$. In particular, the continuum threshold $E_0+1$ is independent of the momentum $\xi$. Throughout the paper, an \emph{excited eigenvalue} means an eigenvalue of $H_g(\xi)$ in $(E_g(\xi),E_0+1)$. For $g=0$, the absence of excited eigenvalues is immediate.

At weak coupling, Seiringer \cite{Seiringer2023} proved that the zero-momentum fiber has
no excited eigenvalue. Earlier, Dahlb\ae k \cite[Theorem 1.3]{Dahlbaek2017} showed that there are $L,g_*>0$ such that for $0<|g|<g_*$, a second discrete eigenvalue can occur only for small momenta $|\xi| \le L |g|$. In the following theorem, we \rev{exclude excited eigenvalues in the region $0<|\xi|\le1/4$}.

\begin{theorem}\label{thm:main}
\rev{There exists $g_*>0$ such that, for
$|g|<g_*$ and $0<|\xi|\le 1/4$,}
\begin{equation}\label{eq:main-conclusion}
 \sigma(H_g(\xi))\cap(-\infty,E_0+1)=\{E_g(\xi)\}.
\end{equation}
The eigenvalue $E_g(\xi)$ is simple.
\end{theorem}

Combining \Cref{thm:main} with the results of  \cite{Dahlbaek2017,Seiringer2023} implies the global absence of excited eigenvalues at sufficiently weak coupling.

\begin{corollary}\label{cor:global}
There exists $g_*>0$ such that, for $|g|<g_*$ and every $\xi\in\R^3$, the operator
$H_g(\xi)$ has at most one eigenvalue below its essential spectrum.
\end{corollary}

A recent result of \cite{Schmidt2026} shows that, for every
coupling strength, the continuum threshold is reached at a
finite critical value $\xi_\ast(g)>0$, in the sense that
$E_g(\xi)=E_0+1$ for all $|\xi|\ge\xi_\ast(g)$.
Consequently, $H_g(\xi)$ has no discrete eigenvalues in
this momentum region. Moreover, \cite{Polzer2023} shows
that the bottom of the spectrum is not an eigenvalue
for $|\xi|>\xi_\ast(g)$.

The absence of excited eigenvalues at weak coupling
is in contrast to the strong-coupling regime $|g|\gg1$,
where excited bound states do occur below the essential spectrum. In fact, the number of discrete energy bands,  within an expanding momentum window around zero, tends to infinity as $|g|$ tends to infinity \cite{MitrouskasSeiringer2023}.\medskip

\noindent \textit{Proof outline}. Following the strategy of
\cite{Seiringer2023}, we first use the Birman--Schwinger principle
to show that any excited eigenvalue must lie within a distance
of order $|g|$ from the continuum threshold. In this energy
interval, the sector with at least two phonons remains uniformly
separated from the relevant energies. Eliminating the vacuum
and higher-phonon sectors leads to an eigenvalue condition
involving a Schur complement on the one-phonon sector.
This operator increases as the energy decreases, so it suffices
to establish its non-negativity at the threshold energy.
In the zero-momentum proof \cite{Seiringer2023}, the
relevant bound is exactly saturated, leaving no positive margin
that could directly be carried over to small nonzero momenta.

To establish positivity of the one-phonon threshold operator
$\widehat{\mathcal O}_\xi$, we show that, after translating
the momentum variable, it admits a factorization of the form
\begin{align}
\widehat{\mathcal O}_\xi
&=|q|S_\xi|q| \qquad \text{with}\qquad
S_\xi=1+A_\xi
+|u_\xi\rangle\langle u_\xi|
-|\varphi_\xi\rangle\langle\varphi_\xi|,
\label{eq:outline}
\end{align}
where $A_\xi$ is a bounded operator of order $g^2$ and
$u_\xi=\varphi_\xi+c_\xi^{1/2}t_\xi$, with $\varphi_\xi \in L^2$, a scalar factor $c_\xi>0$, and
\begin{align}
t_\xi(q)=\frac{g}{|q|\,|q+\xi|}.
\end{align}
Crucially, $t_\xi$ belongs to $L^2$ for $\xi\ne0$,
allowing us to keep the positive rank-one term in
the analysis of $S_\xi$. At zero momentum, the corresponding contribution is dropped before the factorization,
since $t_0$ does not belong to $L^2$.

To prove $S_\xi\ge0$, we use the saturated inequality
from \cite{Seiringer2023}, $1+A_0-|\varphi_0\rangle\langle\varphi_0|\ge0$,
with zero as the bottom of its spectrum.
For $\xi\ne0$, the additional positive rank-one term
in \eqref{eq:outline} allows us to reduce positivity
to a scalar inequality with a positive contribution
close to one. The zero-momentum inequality, together with
quantitative estimates comparing the nonzero- and
zero-momentum quantities, ensures that the potentially
negative correction is of order $g^2|\xi|+g^4$.
The resulting inequality is therefore strictly satisfied
at sufficiently weak coupling, uniformly for
$0<|\xi|\le1/4$.

\section{The one-phonon Schur complement}\label{sec:onephonon}

We shall repeatedly use the Lieb--Yamazaki estimate
\begin{equation}\label{eq:form-bound}
 \pm\Phiop(v_g)
 \le \theta\bigl((\Pf-p)^2+N\bigr)
     +C\frac{g^2}{\theta},
 \qquad p\in\R^3,\quad 0< \theta \le 1 ,\quad  |g|\le\theta ,
\end{equation}
where the constant $C>0$ is independent of $p$, $g$,  and $\theta$; see
\Cref{app:weighted}. In particular, all fibers have
the common quadratic form domain
\begin{align}
 \mathcal Q=D(|\Pf|)\cap D(N^{1/2}).
\end{align}

Let $\Omega$ denote the normalized Fock space vacuum, and set
\begin{align}
 \Pi^n=\one_{\{N=n\}},
 \qquad
 \Pi^{\ge n}=\one_{\{N\ge n\}},
 \qquad n\in\mathbb N_0.
\end{align}
Identifying the one-phonon sector $\Pi^1\F$ with $L^2(\R^3)$, we write
\begin{align}
 \mathcal Q_1:=D(|k|)\simeq\Pi^1\mathcal Q.
\end{align}
Equivalently, for any fixed $\xi\in\R^3$, this is the
space of all $f$ such that
$(1+|k-\xi|^2)^{1/2}f\in L^2(\R^3)$.

We further set
\begin{equation}\label{eq:nu-def}
 \nu_n(\xi):=
 \inf_{\substack{\Psi\in \Pi^{\ge n}\mathcal Q\\ \|\Psi\|=1}}
 \bigl\langle
 \Psi,\bigl(H_g(\xi)-E_0-1\bigr)\Psi
 \bigr\rangle .
\end{equation}

The next lemma shows that the energy in the sector with at least two
phonons remains uniformly above the threshold $E_0+1$, whereas the
bottom of the nonvacuum sector can lie below this threshold by at most
a distance of order $|g|$.

\begin{lemma}\label{lem:sector-bounds}
\rev{There are constants $g_*>0$ and $C>0$ such that, for
$|\xi|\le 1/4$ and $0<|g|\le g_*$,}
\begin{equation}\label{eq:sector-bounds}
 |E_0|\le \rev{C}g^2,
 \qquad
 \nu_1(\xi)\ge-\rev{C}|g|,
 \qquad
 \nu_2(\xi)\ge\frac14.
\end{equation}
Moreover,
\begin{equation}\label{eq:variational-upper}
 E_g(\xi)\le E_0+|\xi|^2.
\end{equation}
\end{lemma}

\begin{proof}
The vacuum trial state gives $E_0\le0$. Taking $p=0$ and a fixed sufficiently
small $\theta$ in \eqref{eq:form-bound} gives $E_0\ge-Cg^2$. Likewise,
\begin{equation}\label{eq:sector-lower-preliminary}
 H_g(\xi)-E_0-1
 \ge(1-\theta)\bigl((\Pf-\xi)^2+N\bigr)
 -1-E_0-Cg^2/\theta.
\end{equation}
On $\Pi^{\ge1}\F$, for $0<|g|\le \rev{g_*}$, choose $\theta=|g|$ and use
$N\ge1$ to obtain
$\nu_1(\xi)\ge-\rev{C}|g|$.  On
$\Pi^{\ge2}\F$, choose, for example, $\theta=1/4$. Then
\begin{align}
 \Pi^{\ge2}(H_g(\xi)-E_0-1)\Pi^{\ge2}
 \ge\bigl(\tfrac12-E_0-Cg^2\bigr)\Pi^{\ge2},
\end{align}
which is bounded below by $1/4$ for sufficiently small $|g|$.

By \eqref{eq:threshold}, $E_0$ is an isolated eigenvalue of $H_g(0)$. Let
$\Psi_0$ be its normalized ground state. The ground state is unique by
positivity improvement \cite{Miyao2010}. Rotational invariance then implies
$\langle\Psi_0,\Pf\Psi_0\rangle=0$. Since
\[
 H_g(\xi)=H_g(0)-2\xi\cdot\Pf+|\xi|^2,
\]
the variational principle gives \eqref{eq:variational-upper}.
\end{proof}

\noindent \textit{Localization of an excited eigenvalue.}  For $|\xi|\le \rev{1/4}$ and sufficiently small $|g|$,
\eqref{eq:variational-upper} gives $E_g(\xi)<E_0+1$. Thus the ground state
exists in the momentum range considered in \Cref{thm:main}. Its eigenvalue is simple
and the ground state eigenvector has a nonzero vacuum component \cite{Miyao2010,Polzer2023}.

Since $E_0$ is the global minimum of the energy-momentum relation
\cite{LampartMitrouskasMysliwy2023,Polzer2023}, every eigenvalue below the essential
spectrum can be written as
\begin{equation}\label{eq:eigenvalue-parameter}
 E_0+1-\eps,
 \qquad 0<\eps\le1,
\end{equation}
where $\eps$ denotes the distance to the threshold. Suppose that $E_0+1-\eps$ is an excited eigenvalue. If
$\eps>-\nu_1(\xi)$, then the compression
$\Pi^{\ge1}(H_g(\xi)-E_0-1+\eps)\Pi^{\ge1}$ has a bounded inverse
on $\operatorname{Ran}\Pi^{\ge1}$. The Birman--Schwinger principle, applied with respect to the
vacuum sector, then gives
\begin{equation}\label{eq:scalar-epsilon}
 \eps-(1+E_0-|\xi|^2)
 =\big\langle v_g,
 \big[\Pi^{\ge1}(H_g(\xi)-E_0-1+\eps)\Pi^{\ge1}\big]^{-1}
 v_g\big\rangle.
\end{equation}
The left-hand side is strictly increasing in $\eps$, whereas the right-hand
side is nonincreasing. Thus \eqref{eq:scalar-epsilon} has at most one solution
in $(-\nu_1(\xi),\infty)$. That solution corresponds to the ground state energy,
as \eqref{eq:variational-upper} implies
\begin{align}
 E_0+1-E_g(\xi)\ge1-|\xi|^2>-\nu_1(\xi)
\end{align}
for sufficiently small $|g|$. Consequently, the parameter $\eps$ associated with any excited eigenvalue satisfies $\eps \le  -\nu_1(\xi)$. We summarize the above in the following lemma.

\begin{lemma}\label{lem:eps-localization} \rev{There are constants $g_*>0$ and $C>0$ such that, for $0<|g|\le g_*$ and $|\xi|\le 1/4$,} every excited eigenvalue $E_0+1-\eps$ of $H_g(\xi)$ satisfies
\begin{equation}\label{eq:eps-localization} 0<\eps\le-\nu_1(\xi)\le \rev{C}|g|.
\end{equation}
\end{lemma}

\noindent\textit{The one-phonon Schur complement.} In the following, all quadratic forms and resolvents are considered for $0\le \eps \le \rev{C} |g|$. By the previous lemma, every possible excited eigenvalue
has the form $E_0+1-\eps$ with $0<\eps\le \rev{C}|g|$. For sufficiently small $|g|$, we have
\begin{equation}\label{eq:denominator-lower}
1+ E_0-|\xi|^2-\eps\ge\frac12,
\end{equation}
while \Cref{lem:sector-bounds} controls the block with at least two
phonons. In particular,
\begin{equation}\label{eq:Xeps}
X_\xi^{(\eps)}=
\left[\Pi^{\ge2}(H_g(\xi)-E_0-1+\eps)\Pi^{\ge2}\right]^{-1}
\end{equation}
is a bounded nonnegative operator on $\Pi^{\ge2}\mathcal F$ satisfying
\begin{equation}\label{eq:Xeps-bound}
\norm{X_\xi^{(\eps)}}\le4.
\end{equation}
Thus we can define the one-phonon Schur complement
\begin{equation}\label{eq:Odef}
\mathcal O_\xi^{(\eps)}
=\eps+(k-\xi)^2-E_0-D_\xi^{(\eps)}
+\frac{|v_g\rangle\langle v_g|}{1+E_0-|\xi|^2-\eps},
\end{equation}
understood as a quadratic form on $\mathcal Q_1$, where
\begin{equation}\label{eq:Deps}
D_\xi^{(\eps)}
=\Pi^1\Phiop(v_g)\Pi^{\ge2}X_\xi^{(\eps)}
\Pi^{\ge2}\Phiop(v_g)\Pi^1.
\end{equation}
By \eqref{eq:form-bound} and \eqref{eq:Xeps-bound}, the last two terms in \eqref{eq:Odef} are relatively form-bounded with respect to the free part, with sufficiently small relative bounds. Hence $\mathcal O_\xi^{(\eps)}$ defines a semibounded closed quadratic form on $\mathcal Q_1$.

We now apply the Birman--Schwinger principle. This implies that $E_0+1-\eps$ is an eigenvalue of $H_g(\xi)$ if and only if
$0$ is an eigenvalue of the self-adjoint operator associated with
$\mathcal O_\xi^{(\eps)}$. Since
$0\le D_\xi^{(\eps)}\le D_\xi^{(0)}$ and $(1+E_0-|\xi|^2-\eps)^{-1}\ge (1+E_0-|\xi|^2)^{-1}$, we further have
\begin{equation}\label{eq:Omonotonicity}
\mathcal O_\xi^{(\eps)}\ge\eps+\mathcal O_\xi^{(0)}.
\end{equation}
We call $\mathcal O_\xi^{(0)}$ the one-phonon threshold form. To exclude the existence of an excited eigenvalue, it is therefore sufficient
to prove
\begin{equation}\label{eq:threshold-positivity-goal}
\mathcal O_\xi^{(0)}\ge0 \qquad \text{for all} \qquad
0<|\xi|\le \rev{1/4}.
\end{equation}
Indeed, since $\eps>0$, this contradicts the preceding Birman--Schwinger eigenvalue condition that zero must be an eigenvalue of $\mathcal O_\xi^{(\eps)}$.

\section{The threshold form and proof of the main result}
\label{sec:threshold-form}

We first derive a suitable representation of the threshold form  \eqref{eq:factorization}. We then recall the zero-momentum identity from \cite{Seiringer2023}, and state some estimates in \Cref{prop:analytic-estimates} that are needed for the comparison with the case $\xi=0$. After establishing a suitable positivity criterion, we combine these ingredients to conclude the proof of the main result.\medskip

\noindent \textit{Resolvents and the kernel representation.} For $n\in\{0,1,2\}$ and $p\in\R^3$, we define
\begin{align}
R_p^{(n)}
:=
\Pi^{\ge n}
\left[
\Pi^{\ge n}
\bigl(H_g(-p) -E_0+1-n\bigr)
\Pi^{\ge n}
\right]^{-1}
\Pi^{\ge n}.
\end{align}
Following the notation of \cite[Section~2]{Seiringer2023}, we set
\begin{align}
Z_p=R_p^{(0)},\qquad
Y_p=R_p^{(1)},\qquad
X_p=R_{-p}^{(2)}.
\end{align}
For
sufficiently small $|g|$, one readily verifies
\begin{equation}\label{eq:YZ-uniform}
 0\le Y_p\le4,
 \qquad 0\le Z_p\le1,
 \qquad 0\le X_p\le4,
 \qquad p\in\R^3.
\end{equation}
The bound for $Z_p$ follows from
$\inf\sigma(H_g(-p))\ge E_0$, see
\cite{LampartMitrouskasMysliwy2023,Polzer2023}. The other two bounds follow from \eqref{eq:form-bound}. We also introduce
\begin{equation}\label{eq:hp-up}
 h_p=Y_pv_g,
 \qquad
 \mathcal E(p)=-\langle v_g,h_p\rangle,
\end{equation}
and note that
\begin{equation}\label{eq:calE-zero}
 \mathcal E(0)=E_0.
\end{equation}
The latter follows by taking the vacuum Schur complement for $H_g(0)-E_0$.

The next lemma provides a representation of the threshold
form with translated one-phonon momentum, so that the kinetic term becomes $|q|^2$. The
singularity of the form factor then lies at $-\xi$.
The lemma is the nonzero-momentum counterpart of
\cite[Eqns.~(2.4)--(2.5)]{Seiringer2023}.

\begin{lemma}\label{lem:kernel-representation}
\rev{Assume $|g|$ is sufficiently small and $|\xi|\le1/4$.} Let $(\tau_\xi f)(k)=f(k+\xi)$ and
$\widehat{\mathcal O}_\xi
=\tau_\xi\mathcal O_\xi^{(0)}\tau_\xi^{-1}$.
Then, for all $f\in\mathcal Q_1$,
\begin{equation}\label{eq:kernel-form}
 \begin{aligned}
 \langle f,\widehat{\mathcal O}_\xi f\rangle
 ={}&\int_{\R^3}\bigl(|q|^2+\mathcal E(q)-E_0\bigr)|f(q)|^2\,\dd q\\
 &+\iint_{\R^3\times\R^3}
 \overline{f(q)}\frac{g}{|q+\xi|}
 C_\xi(q,r)\frac{g}{|r+\xi|}f(r)\,\dd q\,\dd r,
 \end{aligned}
\end{equation}
where
\begin{equation}\label{eq:Ckernel}
 \begin{aligned}
 C_\xi(q,r)={}&\frac1{1+E_0-|\xi|^2}
 -\langle\Omega-h_q,Z_{\xi+q+r}(\Omega-h_r)\rangle
 -\langle h_q,X_\xi h_r\rangle.
 \end{aligned}
\end{equation}
\end{lemma}

\begin{proof}
We note the pull-through formulas
\begin{align}
 X_\xi a^*_{\xi+q}
 &=a^*_{\xi+q}Y_q-\frac{g}{|q+\xi|}X_\xi Y_q,
 \label{eq:pullthrough}\\
 a_{\xi+r}Y_q
 &=Z_{\xi+q+r}a_{\xi+r}
 -\frac{g}{|r+\xi|}Z_{\xi+q+r}Y_q.
 \label{eq:pullthrough2}
\end{align}
These identities correspond to \cite[Eqns.~(2.2)--(2.3)]{Seiringer2023} and can be derived in the same way. Inserting them into $D_\xi^{(0)}$ and using the canonical commutation
relations gives, after translation to the $q$ variable, the kernel contribution for $\tau_\xi D_\xi^{(0)} \tau_\xi^{-1}$,
\begin{align}
 \delta(q-r)\langle v_g,Y_qv_g\rangle
 +\frac{g^2}{|q+\xi|\,|r+\xi|}
 \Bigl(\langle\Omega-h_q,Z_{\xi+q+r}(\Omega-h_r)\rangle
 +\langle h_q,X_\xi h_r\rangle\Bigr).
\end{align}
Thus the diagonal contribution is $\langle v_g,Y_qv_g\rangle$, while the two off-diagonal matrix elements are precisely those in \eqref{eq:Ckernel}. Substitution into \eqref{eq:Odef} at $\eps=0$
yields \eqref{eq:kernel-form}.
\end{proof}

The diagonal term in \eqref{eq:kernel-form} involving $\mathcal E(q)-E_0$  is small relative to $|q|^2$. Following the  argument in \cite[Appendix~A]{Seiringer2023}, we have 
\begin{equation}\label{eq:calE-Hessian}
 \sup_{p\in\R^3}\|D^2\mathcal E(p)\|\le Cg^2.
\end{equation}
This follows from the weighted resolvent estimates in \Cref{lem:weighted-resolvents}, see \eqref{eq:self-energy-Hessian-proof} below. Rotational invariance gives $\nabla\mathcal E(0)=0$, and therefore
\begin{equation}\label{eq:calE-quadratic}
 |\mathcal E(q)-E_0|\le Cg^2|q|^2.
\end{equation}

\noindent \textit{Completing the square.} We now decompose the kernel in \eqref{eq:Ckernel} as
\begin{equation}\label{eq:Cdecomposition}
 C_\xi(q,r)=c_\xi+\psi_\xi(q)+\psi_\xi(r)+F_\xi(q,r),
\end{equation}
where
\begin{align}
 c_\xi&=C_\xi(0,0)
 \qquad \text{and} \qquad 
 \psi_\xi(q)=C_\xi(q,0)-c_\xi.
 \label{eq:c-psi-F}
\end{align}
Thus $\psi_\xi(0)=0$ and $F_\xi(q,0)=F_\xi(0,r)=0$. We also define
\begin{equation}\label{eq:Adef}
 A_\xi=\frac{\mathcal E(q)-E_0}{|q|^2}+g^2K_\xi,
\end{equation}
where the first term is a multiplication operator. Since
$\mathcal E$ is $C^2$ and rotation invariant, there is a
scalar $a_g$ such that $
 \mathcal E(q)-E_0=a_g|q|^2+o(|q|^2)$ as $|q| \to 0$.
 The quotient therefore extends continuously to $q=0$,
with value $a_g$. The operator $K_\xi$ is the integral operator
with kernel
\begin{equation}\label{eq:Kdef}
 K_\xi(q,r)=
 \frac{F_\xi(q,r)}
 {|q|\,|q+\xi|\,|r|\,|r+\xi|}.
\end{equation}
By \eqref{eq:calE-quadratic}, the multiplication operator in
\eqref{eq:Adef} has norm $O(g^2)$. For $\xi=0$, the operator $1+A_0$
coincides with the operator $1+A$ in \cite{Seiringer2023}.

For $g\ne0$, we further set
\begin{equation}\label{eq:b-t-def}
 t_\xi(q)=\frac{g}{|q|\,|q+\xi|},
 \qquad
 b_\xi(q)=\frac{g\psi_\xi(q)}{|q|\,|q+\xi|},
 \qquad
 \varphi_\xi=c_\xi^{-1/2}b_\xi,
\end{equation}
and note that
\begin{equation}\label{eq:Riesz}
 \norm{t_\xi}_2^2
 =g^2\int_{\R^3}\frac{\dd q}{|q|^2|q+\xi|^2}
 =\frac{\pi^3g^2}{|\xi|}.
\end{equation}
Here the last identity follows, for example, from
\begin{align}
 \frac1{|q|^2|q+\xi|^2}
 &=\int_0^1
 \frac{\dd t}
 {\bigl(|q+(1-t)\xi|^2+t(1-t)|\xi|^2\bigr)^2},
\end{align}
together with the elementary formula
$\int_{\R^3} (|q|^2+b)^{-2}\,\dd q=\pi^2/\sqrt b$ for $b>0$.

Thus we have $t_\xi\in L^2(\R^3)$ for $\xi\ne0$. \Cref{prop:analytic-estimates}
below shows that $b_\xi\in L^2(\mathbb R^3)$, $1+A_\xi\ge\frac12$,
and $c_\xi\ge\rev{\kappa}(g^2+|\xi|^2)$. In particular, $c_\xi>0$ whenever
$(g,\xi)\ne(0,0)$ in the parameter range under consideration.

For $0<|\xi|\le \rev{1/4}$, we can complete the square in
\eqref{eq:kernel-form}, so that
\begin{equation}\label{eq:factorization}
 \widehat{\mathcal O}_\xi=|q|S_\xi|q|,
 \qquad
 S_\xi=1+A_\xi
 +|\varphi_\xi+c_\xi^{1/2}t_\xi\rangle
  \langle\varphi_\xi+c_\xi^{1/2}t_\xi|
 -|\varphi_\xi\rangle\langle\varphi_\xi|.
\end{equation}
Indeed, the constant and one-variable parts of
\eqref{eq:Cdecomposition} give
\begin{align}
 c_\xi|t_\xi\rangle\langle t_\xi|
 +|b_\xi\rangle\langle t_\xi|
 +|t_\xi\rangle\langle b_\xi|,
\end{align}
which is precisely the difference of the two rank-one terms in \eqref{eq:factorization}.\medskip

\noindent \textit{The $\xi=0$ case.} At zero momentum, the decomposition underlying \eqref{eq:factorization} coincides with the one used in \cite{Seiringer2023}, with the crucial difference that $t_0(q)=g/|q|^2$ does not belong to $L^2(\R^3)$. In \cite{Seiringer2023}, the corresponding positive rank-one term in the one-phonon form is therefore dropped before the factorization. This gives the lower bound
\begin{align}
\widehat{\mathcal O}_0 \ge |q| \, \mathsf S\, |q|, \qquad \mathsf S : =   1+A_0-|\varphi_0\rangle\langle\varphi_0|.
\end{align}

As explained in \cite[Remark~2.2]{Seiringer2023}, the discarded positive term cannot be used to obtain a stronger lower bound when $\xi=0$. Thus, $ \mathsf S$ is the relevant operator at zero momentum.

Setting $\beta_0:=\langle\varphi_0,(1+A_0)^{-1}\varphi_0\rangle$, the Birman--Schwinger criterion gives $ \mathsf S \ge0$ if and only if $\beta_0\le1$, with $\ker  \mathsf S \neq\{0\}$ precisely when $\beta_0=1$. The positivity criterion is exactly saturated as expressed by the following identity from \cite[Lemma~2.1]{Seiringer2023}.

\begin{lemma}[Zero-momentum identity]\label{lem:seiringer-identity}
For all sufficiently small $g\ne0$,
\begin{equation}\label{eq:zero-anchor}
\langle\varphi_0,(1+A_0)^{-1}\varphi_0\rangle=1.
\end{equation}
\end{lemma}

This identity is non-perturbative in the following sense: The Birman--Schwinger criterion is exactly saturated at one for all sufficiently small $g$. A finite-order expansion of $(1+A_0)^{-1}$ with a remainder of uncontrolled sign therefore cannot decide the required inequality, even though $A_0=O(g^2)$.

For $\xi\ne0$, we keep the positive rank-one term in  \eqref{eq:factorization}. In fact, this term provides a crucial positive contribution used in the proof below.\medskip

\noindent \textit{Positivity criterion for $S_\xi$.} For $\xi \neq 0$, we set
\begin{equation}\label{eq:NTU}
 N_\xi=\langle b_\xi,(1+A_\xi)^{-1}b_\xi\rangle,
 \quad
 U_\xi=\langle b_\xi,(1+A_\xi)^{-1}t_\xi\rangle,
 \quad
 T_\xi=\langle t_\xi,(1+A_\xi)^{-1}t_\xi\rangle.
\end{equation}
The definition of $N_\xi$ also applies at $\xi=0$, unlike those of
$U_\xi$ and $T_\xi$. The estimates below provide bounds on $U_\xi$ and $T_\xi$ and quantify the deviations of $c_\xi$ and $N_\xi$ from their zero-momentum counterparts.
 The proposition is proved in
Section~\ref{sec:estimates}.

\begin{proposition}\label{prop:analytic-estimates}
\rev{There exist $g_*>0$, $C>0$, and $\kappa>0$ such that,
for $0<|g|<g_*$ and $|\xi|\le1/4$,} $K_\xi$ is Hilbert--Schmidt,
$b_\xi\in L^2$, and
\begin{equation}\label{eq:analytic-Ac}
 \norm{A_\xi}\le \rev{C}g^2,
 \qquad
 1+A_\xi\ge\frac12,
 \qquad
 c_\xi\ge\rev{\kappa}(g^2+|\xi|^2).
\end{equation}
Moreover,
\begin{align}\label{eq:analytic-UT}
 |U_\xi|&\le \rev{C}g^2,
 \quad
 0\le T_\xi\le \rev{C}\frac{g^2}{|\xi|},
 \quad
 |c_\xi-c_0|\le \rev{C}|\xi|^2,
 \quad
 |N_\xi-N_0|\le \rev{C}g^2|\xi|
\end{align}
for all $0<|\xi|\le \rev{1/4}$.
\end{proposition}

Introducing the positive invertible operator\begin{align}
G_\xi
=1+A_\xi
+|\varphi_\xi+c_\xi^{1/2}t_\xi\rangle
\langle\varphi_\xi+c_\xi^{1/2}t_\xi|,
\end{align}
we have
\begin{align}
S_\xi\ge0
\quad\Longleftrightarrow\quad
\langle\varphi_\xi,G_\xi^{-1}\varphi_\xi\rangle\le1.
\end{align}
The rank-one (Sherman--Morrison) inversion  formula gives
\begin{align}
G_\xi^{-1}=(1+A_\xi)^{-1}
-\frac{(1+A_\xi)^{-1}
|\varphi_\xi+c_\xi^{1/2}t_\xi\rangle
\langle\varphi_\xi+c_\xi^{1/2}t_\xi|
(1+A_\xi)^{-1}}
{1+\langle\varphi_\xi+c_\xi^{1/2}t_\xi,
(1+A_\xi)^{-1}
(\varphi_\xi+c_\xi^{1/2}t_\xi)\rangle}.
\end{align}
Substituting \eqref{eq:NTU} and simplifying yields
\begin{equation}\label{eq:rank-one-scalar-identity}
1-\langle\varphi_\xi,G_\xi^{-1}\varphi_\xi\rangle
=\frac{|1+U_\xi|^2+T_\xi(c_\xi-N_\xi)}
{1+\langle\varphi_\xi+c_\xi^{1/2}t_\xi,
(1+A_\xi)^{-1}
(\varphi_\xi+c_\xi^{1/2}t_\xi)\rangle}.
\end{equation}
Since the denominator is positive, we arrive at the positivity criterion
\begin{equation}\label{eq:rank-one-criterion}
S_\xi\ge0
\quad\Longleftrightarrow\quad
|1+U_\xi|^2+T_\xi(c_\xi-N_\xi)\ge0.
\end{equation}
The advantage of this criterion is that \Cref{lem:seiringer-identity} gives $N_0=c_0$. Hence the potentially negative contribution $T_\xi(c_\xi-N_\xi)$ is determined by the deviations of $c_\xi$ and $N_\xi$ from their zero-momentum values. Together with the bound on $T_\xi$ and the fact that $U_\xi$ is small, this allows us to verify the criterion.

\begin{proof}[Proof of \Cref{thm:main}]
Fix \rev{$0<|\xi|\le 1/4$}. By
\Cref{lem:seiringer-identity}, we have
\begin{align}
 N_0=c_0.
\end{align}
Hence
\begin{align}\label{eq:cN-perturbation}
 |c_\xi-N_\xi|
 &\le |c_\xi-c_0|+|N_0-N_\xi|  \le \rev{C}\bigl(|\xi|^2+g^2|\xi|\bigr),
\end{align}
where we used \Cref{prop:analytic-estimates} in the last step.
Together with \eqref{eq:analytic-UT}, this gives
\begin{align}
 |1+U_\xi|^2+T_\xi(c_\xi-N_\xi)
 &\ge(1-|U_\xi|)^2-T_\xi|c_\xi-N_\xi|\notag \\
 &\ge(1-\rev{C}g^2)^2
 -\rev{C}\frac{g^2}{|\xi|}
   \bigl(|\xi|^2+g^2|\xi|\bigr)\notag \\
 &=(1-\rev{C}g^2)^2-\rev{C}\bigl(g^2|\xi|+g^4\bigr) >0
\end{align}
for all sufficiently small $|g|$, uniformly over $0<|\xi|\le \rev{1/4}$.

It follows from \eqref{eq:rank-one-criterion} that $S_\xi\ge0$, and hence,
by \eqref{eq:factorization},
$\mathcal O_\xi^{(0)}\ge0$. The Birman--Schwinger argument from
\Cref{sec:onephonon} thus excludes every excited eigenvalue below
$E_0+1$. Existence of the ground state eigenvalue in the considered momentum range follows from \eqref{eq:variational-upper}. As stated in \Cref{sec:onephonon}, simplicity of the ground state, when it exists, is a known result \cite{Miyao2010,Polzer2023}.
\end{proof}

\begin{remark}[Order of limits]
For $\xi\ne0$, let
$\delta_g(\xi):=1-\langle\varphi_\xi,G_\xi^{-1}\varphi_\xi\rangle$.
Although the numerator in \eqref{eq:rank-one-scalar-identity}
is uniformly positive for sufficiently small $|g|$,
the resulting positivity margin $\delta_g(\xi)$
is not uniform as $\xi\to0$ for fixed $g\ne0$.
Indeed, the denominator in \eqref{eq:rank-one-scalar-identity}
equals
\begin{align*}
1+\frac{N_\xi}{c_\xi}
+2\operatorname{Re}U_\xi+c_\xi T_\xi.
\end{align*}
For fixed sufficiently small $g\ne0$, we have
$c_\xi\to c_0>0$ as $\xi\to0$, while
\eqref{eq:Riesz} and \eqref{eq:analytic-Ac} imply $
T_\xi\ge\frac{\pi^3g^2}{2|\xi|}$.
Thus the denominator diverges as $\xi\to0$.
Since the numerator remains bounded by
\eqref{eq:analytic-UT} and \eqref{eq:cN-perturbation},
it follows that $\delta_g(\xi)\to0$.

On the other hand, for every fixed $0<|\xi|\le1/4$,
we have $\delta_g(\xi)\to1$ as $g\to0$,
by \eqref{eq:analytic-Ac} and the bound
$\|b_\xi\|_2=O(|g|)$.
Thus the two limits do not commute:
\begin{align*}
\lim_{g\to0}\lim_{\xi\to0}\delta_g(\xi)&=0,\qquad \qquad
\lim_{\xi\to0}\lim_{g\to0}\delta_g(\xi)=1.
\end{align*}
This is consistent with the exact saturation of the
zero-momentum Birman--Schwinger bound from \Cref{lem:seiringer-identity}.
\end{remark}

\section{Proof of Proposition \ref{prop:analytic-estimates}}\label{sec:estimates}

Throughout this section, $|g|>0$ is sufficiently small and \rev{$|\xi|\le1/4$}. Unless indicated otherwise, constants \rev{are independent of $g$ and $\xi$}. We first establish the resolvent bounds and the estimates for $c_\xi$. We then prove the bounds on the kernels $\psi_\xi$, $F_\xi$ and $C_\xi$, and derive the required estimates for the quantities $N_\xi$, $U_\xi$ and $T_\xi$. 

\subsection{Resolvents and momentum derivatives}\label{subsec:resolvents}

We recall the resolvent definition from the previous section,
\begin{align}
R_p^{(n)}
=
\Pi^{\ge n}
\left[
\Pi^{\ge n}
\bigl(H_g(-p) -E_0+1-n\bigr)
\Pi^{\ge n}
\right]^{-1}
\Pi^{\ge n},
\end{align}
and denote by
$R_{p,\mathrm{fr}}^{(n)}$ the corresponding free resolvent,
\begin{align}
R_{p,\mathrm{fr}}^{(n)}
&:=
\Pi^{\ge n}
\left[
\Pi^{\ge n}
\bigl((\Pf+p)^2+N-E_0+1-n\bigr)
\Pi^{\ge n}
\right]^{-1}
\Pi^{\ge n}.
\end{align}
We  further set $G_p^{(n)}=(R_{p,\mathrm{fr}}^{(n)})^{1/2}$ and recall the definition $E_0=E_g(0)$.

\begin{lemma}\label{lem:weighted-resolvents}
Let $n\in\{0,1,2\}$. We have, uniformly in $p\in\R^3$,
\begin{align}
\norm{G_p^{(n)}\Phiop(v_g)G_p^{(n)}}
&\le C|g|.
\label{eq:weighted-interaction}
\end{align}
For every multi-index $\alpha$ with $|\alpha|\le3$, uniformly in
$p\in\R^3$,
\begin{align}
\norm{(G_p^{(n)})^{-1}
(\partial_p^\alpha R_p^{(n)})
(G_p^{(n)})^{-1}}
&\le C_\alpha,
\label{eq:weighted-derivative-bound}\\
\norm{(G_p^{(n)})^{-1}
\partial_p^\alpha
\bigl(R_p^{(n)}-R_{p,\mathrm{fr}}^{(n)}\bigr)
(G_p^{(n)})^{-1}}
&\le C_\alpha |g|,
\label{eq:resolvent-first-order}\\
\norm{(G_p^{(n)})^{-1}
\partial_p^\alpha
\bigl(
R_p^{(n)}-R_{p,\mathrm{fr}}^{(n)}
+R_{p,\mathrm{fr}}^{(n)}\Phiop(v_g)R_{p,\mathrm{fr}}^{(n)}
\bigr)
(G_p^{(n)})^{-1}}
&\le C_\alpha g^2.
\label{eq:resolvent-remainder}
\end{align}
Moreover,
\begin{align}
\norm{\partial_p^\alpha h_p}
&\le C_\alpha|g|,
\qquad |\alpha|\le3.
\label{eq:h-derivatives}
\end{align}
\end{lemma}

\begin{proof} Since $N\ge n$
on $\Pi^{\ge n}\F$ and $E_0\le0$, we have
\begin{align}
\bigl(R_{p,\mathrm{fr}}^{(n)}\bigr)^{-1}
&\ge \Pi^{\ge n}\bigl((\Pf+p)^2+1\bigr)\Pi^{\ge n}.
\label{eq:free-resolvent-lower}
\end{align}
Moreover, for $n\in\{0,1,2\}$, we have $
N\le 2\bigl(N-E_0+1-n\bigr)$ on $\Pi^{\ge n}\F$, so $(\Pf+p)^2+N+1$ is bounded above by a constant multiple of $(R_{p,\mathrm{fr}}^{(n)})^{-1}$.

The Lieb--Yamazaki estimate \eqref{eq:form-bound}, with $\theta=|g|$,
therefore gives, for $\Psi\in\Pi^{\ge n}\mathcal Q$,
\begin{align}
\bigl|\langle\Psi,\Phiop(v_g)\Psi\rangle\bigr|
&\le |g|\langle\Psi,\bigl((\Pf+p)^2+N\bigr)\Psi\rangle
+C|g|\norm{\Psi}^2 \le C|g|\langle\Psi,
\bigl(R_{p,\mathrm{fr}}^{(n)}\bigr)^{-1}\Psi\rangle.
\end{align}
This proves \eqref{eq:weighted-interaction}.

In the remainder of the proof, all operators are regarded as acting on the
reduced Hilbert space $\Pi^{\ge n}\F$, and the corresponding projections are
suppressed from the notation. Set
\begin{align}
A_p^{(n)}
&:=G_p^{(n)}\Phiop(v_g)G_p^{(n)},
&
M_p^{(n)}
&:=1+A_p^{(n)}.
\end{align}
By \eqref{eq:weighted-interaction}, $M_p^{(n)}$ is uniformly invertible for
sufficiently small $|g|$, and
\begin{align}
R_p^{(n)}
&=G_p^{(n)}(M_p^{(n)})^{-1}G_p^{(n)}.
\label{eq:resolvent-factorization}
\end{align}

Writing
\begin{align}
G_p^{(n)}
&=\bigl((\Pf+p)^2+B_n\bigr)^{-1/2},
&
B_n
&=N-E_0+1-n,
\end{align}
we have $B_n\ge1$ on $\Pi^{\ge n}\F$. For example,
\begin{align}
(G_p^{(n)})^{-1}\partial_{p_i}G_p^{(n)}
&=-({\Pf}_i+p_i)\bigl((\Pf+p)^2+B_n\bigr)^{-1}.
\end{align}
Repeated differentiation therefore yields
\begin{align}
\norm{(G_p^{(n)})^{-1}\partial_p^\alpha G_p^{(n)}}
&\le C_\alpha,
\qquad |\alpha|\le3,
\label{eq:G-derivatives}
\end{align}
uniformly in $p\in\R^3$ and $n\in\{0,1,2\}$.

Together with \eqref{eq:weighted-interaction}, these bounds and the product
rule imply
\begin{align}
\norm{\partial_p^\alpha A_p^{(n)}}
&\le C_\alpha|g|,
\qquad |\alpha|\le3.
\label{eq:A-derivatives}
\end{align}
Indeed, every term produced by the product rule is of the form
\begin{align}
(\partial_p^\beta G_p^{(n)})\Phiop(v_g)
(\partial_p^\gamma G_p^{(n)}),
\qquad \beta+\gamma=\alpha,
\end{align}
and can be estimated by inserting $(G_p^{(n)})^{-1}G_p^{(n)}$ on either
side of $\Phiop(v_g)$ and using \eqref{eq:G-derivatives} and
\eqref{eq:weighted-interaction}.

Since $M_p^{(n)}$ is uniformly invertible for sufficiently small $|g|$,
repeated differentiation of
\begin{align}
\partial_{p_i}(M_p^{(n)})^{-1}
&=-(M_p^{(n)})^{-1}
(\partial_{p_i}M_p^{(n)})
(M_p^{(n)})^{-1}
\end{align}
and \eqref{eq:A-derivatives} show that
\begin{align}
\norm{\partial_p^\alpha(M_p^{(n)})^{-1}}
&\le C_\alpha,
\qquad |\alpha|\le3.
\label{eq:M-inverse-derivatives}
\end{align}

Differentiating \eqref{eq:resolvent-factorization} and using
\eqref{eq:G-derivatives} and \eqref{eq:M-inverse-derivatives}
proves \eqref{eq:weighted-derivative-bound}. For the two difference
estimates, we use the exact identities
\begin{align}
(M_p^{(n)})^{-1}-1
&=-A_p^{(n)}(M_p^{(n)})^{-1},
\label{eq:inverse-first-order}\\
(M_p^{(n)})^{-1}-1+A_p^{(n)}
&=(A_p^{(n)})^2(M_p^{(n)})^{-1}.
\label{eq:inverse-second-order}
\end{align}
Multiplying these identities by $G_p^{(n)}$ from the left and right gives,
respectively, the resolvent differences appearing in
\eqref{eq:resolvent-first-order} and \eqref{eq:resolvent-remainder}. Since
$A_p^{(n)}$ and all its $p$-derivatives up to order three are $O(|g|)$,
while the corresponding derivatives of $(M_p^{(n)})^{-1}$ are uniformly
bounded, differentiating \eqref{eq:inverse-first-order} and
\eqref{eq:inverse-second-order}, and using \eqref{eq:G-derivatives} for the
outer factors, proves \eqref{eq:resolvent-first-order} and
\eqref{eq:resolvent-remainder}.

We shall also use the elementary bound
\begin{equation}\label{eq:shifted-endpoint}
 \sup_{p\in\R^3}
 \int_{\R^3}\frac{\dd k}{|k|^2(|k+p|^2+b)}
 \le \frac{2\pi^2}{\sqrt b},
 \qquad b>0.
\end{equation}
Indeed, by radial rearrangement the integral on the left is maximal at
$p=0$, and spherical coordinates give
\begin{align}
 \int_{\R^3}\frac{\dd k}{|k|^2(|k|^2+b)}
 =4\pi\int_0^\infty\frac{\dd r}{r^2+b}
 =\frac{2\pi^2}{\sqrt b}.
\end{align}
For $n=1$, taking $b=1-E_0\ge1$ therefore gives
\begin{align}
\norm{G_p^{(1)}v_g}^2
&=\int_{\R^3}\frac{|v_g(k)|^2}{|k+p|^2+1-E_0}\,\dd k
\le Cg^2.
\label{eq:endpoint-Gv}
\end{align}
Since $h_p=Y_pv_g=R_p^{(1)}v_g$, combining
\eqref{eq:weighted-derivative-bound} with \eqref{eq:endpoint-Gv} and the
uniform bound on $G_p^{(1)}$ proves \eqref{eq:h-derivatives}.
\end{proof}

For completeness, the same estimates give the self-energy bound used in
Section~\ref{sec:threshold-form}. In fact, differentiating
$p\mapsto \mathcal E(p)=-\langle v_g,Y_pv_g\rangle$ twice, we obtain
\begin{align}\label{eq:self-energy-Hessian-proof}
|\partial_i\partial_j\mathcal E(p)|
&\le
\norm{G_p^{(1)}v_g}^2
\norm{(G_p^{(1)})^{-1}
(\partial_i\partial_jY_p)
(G_p^{(1)})^{-1}}
\le Cg^2.
\end{align}
This proves \eqref{eq:calE-Hessian}. Rotational invariance gives
$\nabla\mathcal E(0)=0$, and hence \eqref{eq:calE-quadratic}.

\subsection{The coefficient \texorpdfstring{$c_\xi$}{c-xi}}
\label{subsec:cxi}

We next estimate the coefficient $c_\xi=C_\xi(0,0)$. Recalling \eqref{eq:Ckernel}, its definition is
\begin{align}
c_\xi
&=
\frac{1}{1+E_0-|\xi|^2}
-\langle \Omega-h_0,Z_\xi(\Omega-h_0)\rangle
-\langle h_0,X_\xi h_0\rangle.
\label{eq}
\end{align}

\begin{lemma}\label{lem:cxi}
We have
\begin{align}
c_0
&=
\frac{\pi^2g^2}{\sqrt{1-E_0}}+O(g^4),
\label{eq:c0-expansion}
\end{align}
and
\begin{align}
c_\xi-c_0
&=
\frac{|\xi|^2}
{(1+E_0)(1+E_0-|\xi|^2)}
+
\frac{|\xi|^2}
{(1-E_0)(1-E_0+|\xi|^2)}
+
O(g^2|\xi|^2).
\label{eq:c-defect}
\end{align}
In particular, for some $\kappa>0$,
\begin{align}
c_\xi
&\ge
\kappa\bigl(g^2+|\xi|^2\bigr),
&
|c_\xi-c_0|
&\le
C|\xi|^2.
\label{eq:c-positive}
\end{align}
\end{lemma}
\begin{proof}
Since $h_0=Y_0v_g=R_0^{(1)}v_g$, \eqref{eq:resolvent-remainder}
with $n=1$ and $p=0$ gives
\begin{align}
h_0
&=
R_{0,\mathrm{fr}}^{(1)}v_g
-
R_{0,\mathrm{fr}}^{(1)}
\Phiop(v_g)
R_{0,\mathrm{fr}}^{(1)}v_g
+r_g.
\label{eq:h0:expansion}
\end{align}
By \eqref{eq:resolvent-remainder} and \eqref{eq:endpoint-Gv}, the
remainder satisfies
\begin{align}
\norm{(G_0^{(1)})^{-1}r_g}
&\le
\left\|
(G_0^{(1)})^{-1}
\left(
R_0^{(1)}
-R_{0,\mathrm{fr}}^{(1)}
+R_{0,\mathrm{fr}}^{(1)}
\Phiop(v_g)
R_{0,\mathrm{fr}}^{(1)}
\right)
(G_0^{(1)})^{-1}
\right\|
\norm{G_0^{(1)}v_g}
\notag\\
&\le
C|g|^3.
\label{eq:h0:remainder}
\end{align}
Since $\norm{G_0^{(1)}}\le 1$, this also gives
$\norm{r_g}\le C|g|^3$.

The first term on the right-hand side of \eqref{eq:h0:expansion}
belongs to the one-phonon sector, whereas the second belongs to the two-phonon
sector, and is therefore orthogonal to the first term. Moreover, by
\eqref{eq:weighted-interaction} and \eqref{eq:endpoint-Gv},
\begin{align}
\left\|
R_{0,\mathrm{fr}}^{(1)}
\Phiop(v_g)
R_{0,\mathrm{fr}}^{(1)}v_g
\right\|
&\le
\norm{G_0^{(1)}}
\norm{G_0^{(1)}\Phiop(v_g)G_0^{(1)}}
\norm{G_0^{(1)}v_g} \le Cg^2.
\label{eq:h0:second-term}
\end{align}
It thus follows from \eqref{eq:h0:expansion}, \eqref{eq:h0:second-term},
and \eqref{eq:h0:remainder} that
\begin{align}
\left\|
h_0-R_{0,\mathrm{fr}}^{(1)}v_g
\right\|
&\le
Cg^2,
\label{eq:h0:correction}
\end{align}
and hence, using \eqref{eq:endpoint-Gv},
\begin{align}
\norm{h_0}
&\le
\norm{R_{0,\mathrm{fr}}^{(1)}v_g}
+Cg^2  \le
\norm{G_0^{(1)}}
\norm{G_0^{(1)}v_g}
+Cg^2
\le
C|g|.
\label{eq:h0:bound}
\end{align}

Since $E_0=-\langle v_g,h_0\rangle$, the one-phonon component of
\eqref{eq:h0:expansion} gives
\begin{align}\label{eq:E0:g4:expansion}
-E_0
&=
\left\langle
v_g,
R_{0,\mathrm{fr}}^{(1)}v_g
\right\rangle
+
\langle v_g,r_g\rangle = 
\left\langle
v_g,
R_{0,\mathrm{fr}}^{(1)}v_g
\right\rangle
+ O(g^4),
\end{align}
where the remainder is estimated in the same weighted sense as above:
\begin{align}
|\langle v_g,r_g\rangle|
\le \|G_0^{(1)}v_g\|\,\|(G_0^{(1)})^{-1}r_g\|
\le Cg^4.
\end{align}
The first two terms in \eqref{eq:h0:expansion} are orthogonal. Using
\eqref{eq:h0:second-term}, $\norm{r_g}\le C|g|^3$, and
$\norm{R_{0,\mathrm{fr}}^{(1)}v_g}=O(|g|)$, we therefore obtain
\begin{align}
\norm{h_0}^2
&=
\norm{R_{0,\mathrm{fr}}^{(1)}v_g}^2
+
O(g^4) = 
\left\langle
v_g,
(R_{0,\mathrm{fr}}^{(1)})^2v_g
\right\rangle
+
O(g^4).
\label{eq:h0:norm-expansion}
\end{align}
We shall also need the higher-phonon component of $h_0$. Since the
first term in \eqref{eq:h0:expansion} lies in the one-phonon sector,
\eqref{eq:h0:second-term} and $\norm{r_g}\le C|g|^3$ give
\begin{align}
\norm{\Pi^{\ge2}h_0}
&\le
\left\|
R_{0,\mathrm{fr}}^{(1)}
\Phiop(v_g)
R_{0,\mathrm{fr}}^{(1)}v_g
\right\|
+\norm{r_g}
\le
Cg^2.
\label{eq:h0:higher-sector}
\end{align}

Using $E_0=-\langle v_g,h_0\rangle$ and the defining equation for $h_0$, we obtain
\begin{align}
(H_g(0)-E_0)(\Omega-h_0)
&=0.
\end{align}
It follows that
$\Omega-h_0$ is the unique ground-state eigenvector with vacuum coefficient one.
Since $Z_0=(H_g(0)-E_0+1)^{-1}$, it further follows that
\begin{align}
Z_0(\Omega-h_0)
&=
\Omega-h_0.
\end{align}
Evaluating \eqref{eq:Ckernel} at $q=r=\xi=0$ and using $\langle\Omega,h_0\rangle=0$, we obtain
\begin{align}
c_0
&=
-\frac{E_0}{1+E_0}
-\norm{h_0}^2
-\left\langle
\Pi^{\ge2}h_0,
X_0\Pi^{\ge2}h_0
\right\rangle.
\label{eq:c0-ground-formula}
\end{align}
By \eqref{eq:h0:higher-sector} and the uniform bound on $X_0$, the last
term is $O(g^4)$. Since $E_0=O(g^2)$, we also have $  E_0 / (1+E_0 )  = E_0+O(g^4)$.
Using \eqref{eq:E0:g4:expansion} and
\eqref{eq:h0:norm-expansion}, it thus follows that
\begin{align}
c_0
= - E_0 - \| h_0 \|^2 + O(g^4) & =
\left\langle
v_g,
\left(
R_{0,\mathrm{fr}}^{(1)}
-(R_{0,\mathrm{fr}}^{(1)})^2
\right)v_g
\right\rangle
+
O(g^4)
\notag\\
&=
g^2
\int_{\R^3}
\frac{|k|^2-E_0}
{|k|^2(|k|^2+1-E_0)^2}
\,\dd k
+
O(g^4)
\notag\\
&=
g^2
\int_{\R^3}
\frac{\dd k}
{(|k|^2+1-E_0)^2}
+
O(g^4),
\label{eq:c0:integral}
\end{align}
where we used that the difference between the two integrals is $O(|E_0|)$, and hence,
after multiplication by $g^2$, contributes $O(g^2|E_0|)=O(g^4)$.
Finally, for $b>0$, we use
\begin{align}
 \int_{\R^3}\frac{\dd k}{(|k|^2+b)^2}
 =\frac{\pi^2}{\sqrt b},
\end{align}
and taking $b=1-E_0$ proves \eqref{eq:c0-expansion}.

To compare $c_\xi$ with $c_0$, define
\begin{align}
d(p)
&=
\left\langle
\Omega-h_0,
Z_p (\Omega-h_0)
\right\rangle
-\frac{1}{1-E_0+|p|^2}
+\langle h_0,X_p\rev{h_0}\rangle.
\label{eq:d-definition}
\end{align}
We claim that, for $i,j\in\{1,2,3\}$,
\begin{align}
\sup_{|p|\le1/4}
| \partial_i\partial_j d(p) |
&\le
Cg^2.
\label{eq:d-Hessian}
\end{align}
The fraction term in \eqref{eq:d-definition} is the vacuum matrix element
of $R_{\rev{p},\mathrm{fr}}^{(0)}$. Moreover,
\begin{align}
\left\langle
\Omega,
R_{\rev{p},\mathrm{fr}}^{(0)}
\Phiop(v_g)
R_{\rev{p},\mathrm{fr}}^{(0)}
\Omega
\right\rangle
&=
0,
\end{align}
since $\Phiop(v_g)$ changes the phonon number by one. Hence
\eqref{eq:resolvent-remainder}, with $n=0$, shows that
\begin{align}
\partial_i\partial_j
\left(
\langle\Omega,Z_{\rev{p}}\Omega\rangle
-
\frac{1}{1-E_0+|\rev{p}|^2}
\right)
&=
O(g^2)
\end{align}
uniformly for $|\rev{p}|\le1/4$.

The free cross terms vanish,
\begin{align}
\langle\Omega,R_{\rev{p},\mathrm{fr}}^{(0)}h_0\rangle
&=
\langle h_0,R_{\rev{p},\mathrm{fr}}^{(0)}\Omega\rangle
=
0,
\end{align}
because $h_0\in\Pi^{\ge1}\F$. The corresponding terms with
$Z_{\rev{p}}-R_{\rev{p},\mathrm{fr}}^{(0)}$, as well as their first two derivatives,
are $O(g^2)$ by \eqref{eq:resolvent-first-order} and
\eqref{eq:h0:bound}. The term  $\langle h_0 , Z_{\rev{p}} h_0\rangle$,
together with its first two derivatives, is $O(g^2)$ by
\eqref{eq:weighted-derivative-bound} and \eqref{eq:h0:bound}.

Finally, since $X_{\rev{p}}$ acts on $\Pi^{\ge2}\F$, only
$\Pi^{\ge2}h_0$ contributes to $\langle h_0,X_{\rev{p}}h_0\rangle$.
Thus \eqref{eq:h0:higher-sector} and
\eqref{eq:weighted-derivative-bound} show that this term and its
first two derivatives are $O(g^4)$. This proves
\eqref{eq:d-Hessian}.

Since $h_0$ is rotation invariant and the resolvents transform covariantly under rotations, $d$ is rotation invariant and hence radial. Therefore $\nabla d(0)=0$, and \eqref{eq:d-Hessian} gives
\begin{align}
|d(\xi)-d(0)|
&\le
Cg^2|\xi|^2.
\label{eq:d-difference}
\end{align}
By definition, 
\begin{align}
c_\xi
&=
\frac{1}{1+E_0-|\xi|^2}
-\frac{1}{1-E_0+|\xi|^2}
-d(\xi).
\label{eq:cxi-via-d}
\end{align}
Subtracting the corresponding identity at $\xi=0$ and using
\eqref{eq:d-difference} gives
\begin{align}
c_\xi-c_0
&=
\frac{|\xi|^2}
{(1+E_0)(1+E_0-|\xi|^2)}
+
\frac{|\xi|^2}
{(1-E_0)(1-E_0+|\xi|^2)}
+
O(g^2|\xi|^2),
\end{align}
which proves \eqref{eq:c-defect}.

Since $|E_0|\le C|g|^2$, the two explicit terms in
\eqref{eq:c-defect} are positive and bounded below by a constant
times $|\xi|^2$ for sufficiently small $|g|$, uniformly for
$|\xi|\le \rev{1/4}$. Together with
\eqref{eq:c0-expansion}, which gives $c_0\ge\kappa g^2$ for sufficiently
small $|g|$, this yields
\begin{align}
c_\xi
&\ge
\kappa\bigl(g^2+|\xi|^2\bigr),
&
|c_\xi-c_0|
&\le
C|\xi|^2,
\end{align}
after decreasing $\kappa$ if necessary. This proves
\eqref{eq:c-positive}.
\end{proof}

\subsection{Kernel estimates}\label{subsec:kernel}

For the remainder of this section, set
\begin{align}
\omega(q)
&=
\min\{|q|,1\}.
\label{eq:omega}
\end{align}

\begin{lemma}\label{lem:centered-kernel}
The kernel $C_\xi$ defined in \eqref{eq:Ckernel} is real and symmetric,
and satisfies
\begin{align}
C_{R\xi}(Rq,Rr)
&=
C_\xi(q,r),
\qquad R\in O(3).
\end{align}
Moreover, uniformly for $|\xi|\le \rev{1/4}$,
\begin{align}
\sup_{q,r}
\left|
\partial_q^\alpha
\partial_r^\beta
\partial_\xi^\gamma
C_\xi(q,r)
\right|
&\le
C,
\qquad
|\alpha|+|\beta|+|\gamma|\le3.
\label{eq:Cderivatives}
\end{align}
The functions $\psi_\xi$ and $F_\xi$ satisfy
\begin{align}
|\psi_\xi(q)|
&\le
C\omega(q)\bigl(\omega(q)+|\xi|\bigr),
\label{eq:psi-pointwise:1}\\
|\psi_\xi(q)-\psi_0(q)|
&\le
C|\xi|\omega(q),
\label{eq:psi-pointwise}\\
|F_\xi(q,r)|
&\le
C\omega(q)\omega(r),
\label{eq:Fbasic:1}\\
|F_\xi(q,r)-F_0(q,r)|
&\le
C|\xi|\omega(q)\omega(r),
\label{eq:Fbasic}\\
|\nabla_qF_\xi(q,r)|
+
|D_q^2F_\xi(q,r)|
&\le
C\omega(r).
\label{eq:F-derivative-bounds}
\end{align}
\end{lemma}

\begin{proof}
We first prove \eqref{eq:Cderivatives}. Recall from \eqref{eq:Ckernel}
that $C_\xi(q,r)$ is expressed in terms of $h_q$, $h_r$,
$Z_{\xi+q+r}$, and $X_\xi$. By \Cref{lem:weighted-resolvents}, the
resolvents and their momentum derivatives up to order three are
uniformly bounded in the weighted operator norm, while
\eqref{eq:h-derivatives} gives
\begin{align}
\norm{\partial_p^\alpha h_p}
&\le
C_\alpha|g|,
\qquad
|\alpha|\le3.
\end{align}
Since $\norm{G_p^{(n)}}\le1$, the weighted bounds of
\Cref{lem:weighted-resolvents} also imply the corresponding ordinary
operator-norm bounds. Moreover, $1+E_0-|\xi|^2$ is bounded away from
zero for $|\xi|\le \rev{1/4}$. Differentiating \eqref{eq:Ckernel}
with respect to $q$, $r$, and $\xi$, and applying the product rule,
therefore gives
\begin{align}
\sup_{q,r}
\left|
\partial_q^\alpha
\partial_r^\beta
\partial_\xi^\gamma
C_\xi(q,r)
\right|
&\le
C,
\qquad
|\alpha|+|\beta|+|\gamma|\le3,
\end{align}
which proves \eqref{eq:Cderivatives}.

Reality of $C_\xi(q,r)$ and the stated rotation covariance follow
directly from \eqref{eq:Ckernel}. Since the resolvents are self-adjoint,
\eqref{eq:Ckernel} also gives $C_\xi(q,r)=C_\xi(r,q)$.

Since $q\mapsto C_0(q,0)$ is radial, we have
$\nabla_q C_0(0,0)=0$. By \eqref{eq:Cderivatives} and the fundamental
theorem of calculus,
$|\nabla_qC_\xi(0,0)-\nabla_qC_0(0,0)|\le C|\xi|$, and therefore
\begin{align}
|\nabla_qC_\xi(0,0)|
&\le
C|\xi|.
\end{align}

Recalling $\psi_\xi(q)=C_\xi(q,0)-C_\xi(0,0)$, Taylor's theorem in $q$,
together with the uniform bound \eqref{eq:Cderivatives}, gives
$\psi_\xi(q)=\nabla_qC_\xi(0,0)\cdot q+O(|q|^2)$, where the remainder
is uniform for $|\xi|\le \rev{1/4}$. Hence
\begin{align}
|\psi_\xi(q)|
&\le
C|q|\bigl(|q|+|\xi|\bigr)
\end{align}
for $|q|\le1$, while boundedness gives the corresponding estimate for
$|q|\ge1$. This proves \eqref{eq:psi-pointwise:1}.

For the second estimate, differentiating
$\psi_\eta(q)=C_\eta(q,0)-C_\eta(0,0)$ with respect to $\eta$ gives
\begin{align}
\nabla_\eta\psi_\eta(q)
&=
\nabla_\eta C_\eta(q,0)
-
\nabla_\eta C_\eta(0,0).
\end{align}
Hence, by \eqref{eq:Cderivatives} and the fundamental theorem of
calculus in $q$,
\begin{align}
|\nabla_\eta\psi_\eta(q)|
&\le
C\omega(q).
\end{align}
Applying the fundamental theorem of calculus in $\eta$ between $0$ and
$\xi$ yields
\begin{align}
|\psi_\xi(q)-\psi_0(q)|
&\le
C|\xi|\omega(q),
\end{align}
which proves \eqref{eq:psi-pointwise}.

The remaining estimates follow by the same argument. For instance, for
$|q|,|r|\le1$, applying the fundamental theorem of calculus successively in
$q$ and $r$ to the definition of $F_\xi$ in \eqref{eq:Cdecomposition}, we obtain
\begin{align}
F_\xi(q,r)
&=
\sum_{i,j=1}^3 q_i r_j
\int_0^1\!\!\int_0^1
\partial_{q_i}\partial_{r_j}
C_\xi(sq,tr)\,\dd s\,\dd t.
\end{align}
Hence, by \eqref{eq:Cderivatives},
\begin{align}
|F_\xi(q,r)|
&\le
C|q||r|
=
C\omega(q)\omega(r).
\end{align}
If one of the variables has modulus at least one, the same estimate follows
by applying the fundamental theorem of calculus only in the other variable
and using the uniform bounds on the first derivatives of $C_\xi$ from
\eqref{eq:Cderivatives}. If both variables have modulus at least one, the
uniform bound on $C_\xi$ itself suffices. Applying the same argument to $F_\xi-F_0$, with one additional
derivative in $\xi$, yields \eqref{eq:Fbasic}. Finally,
differentiating the definition of $F_\xi$ with respect to $q$ and
applying the fundamental theorem of calculus in $r$ gives
\eqref{eq:F-derivative-bounds}. Only derivatives of $C_\xi$ of total
order at most three are used.
\end{proof}

For the following estimates, we set
\begin{align}
\mu_\xi(q)
&=
\frac{1}{|q|^2|q+\xi|^2}.
\label{eq:singular-weight}
\end{align}
In particular, $\mu_0(q)=|q|^{-4}$.

\begin{lemma}\label{lem:singular-integrals}
Uniformly for $|\xi|\le1/4$,
\begin{align}
\int_{\R^3}
\omega(q)\bigl(\omega(q)+|\xi|\bigr)\mu_\xi(q)\,\dd q
&\le
C,
\label{eq:singular-basic}\\
\int_{\R^3}
\omega(q)^3|\mu_\xi(q)-\mu_0(q)|\,\dd q
&\le
C|\xi|.
\label{eq:measure-comparison}
\end{align}
\end{lemma}

\begin{proof}
For $\xi=0$, both estimates follow directly. We therefore assume
$\xi\neq0$ and split the integrals into $|q|\le2|\xi|$ and
$|q|>2|\xi|$.

On the first region, set $q=|\xi|x$ and $e=\xi/|\xi|$. Since
$|\xi|\le1/4$, we have $\omega(q)=|q|$, and hence
\begin{align}
\int_{|q|\le2|\xi|}
\omega(q)\bigl(\omega(q)+|\xi|\bigr)\mu_\xi(q)\,\dd q
&=
|\xi|
\int_{|x|\le2}
\frac{|x|(|x|+1)}
{|x|^2|x+e|^2}\,\dd x
\le
C|\xi|.
\end{align}
The last integral is bounded uniformly for $e\in S^2$. Similarly,
using $|\mu_\xi-\mu_0|\le\mu_\xi+\mu_0$ and the same change of
variables,
\begin{align}
\int_{|q|\le2|\xi|}
\omega(q)^3|\mu_\xi(q)-\mu_0(q)|\,\dd q
&\le
C|\xi|^2.
\end{align}

On the complementary region $|q|>2|\xi|$, we have
\begin{align}
|q+\xi|
&\ge
\frac12|q|,
&
|\mu_\xi(q)-\mu_0(q)|
&\le
C|\xi||q|^{-5}.
\label{eq:measure-pointwise}
\end{align}
Moreover, $\omega(q)+|\xi|\le C\omega(q)$, and therefore
\begin{align}
\int_{|q|>2|\xi|}
\omega(q)\bigl(\omega(q)+|\xi|\bigr)\mu_\xi(q)\,\dd q
&\le
C\int_{\R^3}
\omega(q)^2|q|^{-4}\,\dd q
\le
C.
\end{align}
For the second integral, radial integration gives
\begin{align}
\int_{|q|>2|\xi|}
\omega(q)^3|\mu_\xi(q)-\mu_0(q)|\,\dd q
&\le
C|\xi|
\left(
\int_{2|\xi|}^1\dd r
+
\int_1^\infty r^{-3}\,\dd r
\right)
\le
C|\xi|.
\end{align}
This completes the proof of the lemma.
\end{proof}

We can now record the operator bounds needed below. By
\eqref{eq:Fbasic:1} and \eqref{eq:singular-basic},
\begin{align}
\norm{K_\xi}_{\mathrm{HS}}^2
&=
\int_{\R^3}\int_{\R^3}
|F_\xi(q,r)|^2\mu_\xi(q)\mu_\xi(r)\,\dd q\,\dd r
\le C,
\notag\\
\norm{b_\xi}_2^2
&=
g^2\int_{\R^3}|\psi_\xi(q)|^2\mu_\xi(q)\,\dd q
\le Cg^2.
\label{eq:HS-and-b}
\end{align}
In particular, $K_\xi$ is bounded with $\norm{K_\xi}\le C$. Since
$K_\xi$ is self-adjoint, so is $A_\xi$, and together with
\eqref{eq:calE-quadratic} and \eqref{eq:Adef} this gives
\begin{align}
\norm{A_\xi}
&\le Cg^2,\qquad\qquad
1+A_\xi \ge \frac12
\label{eq:A-technical}
\end{align}
for sufficiently small $|g|$, uniformly for $|\xi|\le \rev{1/4}$.

\subsection{Estimates for \texorpdfstring{$N_\xi$, $U_\xi$, and $T_\xi$}{N-xi, U-xi, and T-xi}}
\label{subsec:NTU}

We define
\begin{align}
m(q)
&=
1+\frac{\mathcal E(q)-E_0}{|q|^2},
\qquad q\neq0,
\label{eq:m-technical}
\end{align}
and extend $m$ continuously to $q=0$. This is possible since
$\mathcal E$ is radial and $C^2$, so its Hessian at the origin is a
scalar matrix. By \eqref{eq:calE-quadratic},
\begin{align}
\norm{m-1}_\infty
&\le
Cg^2.
\end{align}
In particular, for sufficiently small $|g|$,
\begin{align}
m
&\ge
\frac34.
\end{align}

To estimate $N_\xi-N_0$, we factor out the explicit singular weight in $(1+A_\xi)^{-1}b_\xi$ and study the remaining function in a weighted supremum norm. To this end, we introduce the Banach space
\begin{align}
\mathcal B
&=
\left\{
f:\R^3\to\R \text{ measurable}:
f(0)=0,\quad
\norm{f}_{\mathcal B}
:=
\sup_{q\ne0}\frac{|f(q)|}{\omega(q)}
<\infty
\right\}.
\label{eq:Bnorm}
\end{align}
For each $\xi$, we define the integral operator
\begin{align}
(\mathscr L_\xi f)(q)
&=
\int_{\R^3}
F_\xi(q,r) \mu_\xi(r) f(r)\,\dd r.
\label{eq:Lfamily}
\end{align}
By \eqref{eq:Fbasic:1} and \eqref{eq:singular-basic},
$\mathscr L_\xi$ is a bounded operator on $\mathcal B$, uniformly in
$\xi$.

\begin{lemma}\label{lem:NTU-estimates}
For every $|\xi|\le \rev{1/4}$, the equation
\begin{align}
m\eta_\xi+g^2\mathscr L_\xi\eta_\xi
&=
\psi_\xi
\label{eq:eta-equation}
\end{align}
has a unique solution $\eta_\xi\in\mathcal B$. With $A_\xi$
defined in \eqref{eq:Adef}, this solution satisfies
\begin{align}
\bigl[(1+A_\xi)^{-1}b_\xi\bigr](q)
&=
\frac{g\eta_\xi(q)}{|q|\,|q+\xi|}.
\label{eq:x-eta}
\end{align}
Moreover, uniformly for $|\xi|\le \rev{1/4}$, we have
\begin{align}
|\eta_0(q)|
&\le
C\omega(q)^2,\qquad \quad
|\eta_\xi(q)-\eta_0(q)|
\le
C|\xi|\omega(q).
\label{eq:eta-bounds}
\end{align}
Finally, for $0<|\xi|\le \rev{1/4}$, we have
\begin{align}
|U_\xi|
&\le
Cg^2,
&
0\le T_\xi
&\le
C\frac{g^2}{|\xi|},
&
|N_\xi-N_0|
&\le
Cg^2|\xi|.
\label{eq:scalar-bounds}
\end{align}
\end{lemma}

\begin{proof} \textit{Existence and uniqueness.} By \eqref{eq:psi-pointwise:1}, we have $\psi_\xi\in\mathcal B$. Moreover,
$\norm{\mathscr L_\xi}_{\mathcal B\to\mathcal B}\le C$ by
\eqref{eq:Fbasic:1} and \eqref{eq:singular-basic}, while $m\ge3/4$.
Hence multiplication by $m^{-1}$ is bounded on $\mathcal B$, so
$m^{-1}\psi_\xi\in\mathcal B$, and
\begin{align}
\norm{g^2m^{-1}\mathscr L_\xi}_{\mathcal B\to\mathcal B}
&\le
Cg^2<1
\end{align}
for sufficiently small $|g|$. Therefore
$1+g^2m^{-1}\mathscr L_\xi$ is invertible on $\mathcal B$ by a
Neumann series, and \eqref{eq:eta-equation} has a unique solution
$\eta_\xi\in\mathcal B$ with
$\norm{\eta_\xi}_{\mathcal B}\le C$.

For $\xi=0$, the function $m$, the right-hand side $\psi_0$, and the
operator $\mathscr L_0$ are rotation invariant. Hence, by uniqueness of the solution,
$\eta_0$ is radial.\medskip

\noindent\emph{Quadratic vanishing at $\xi=0$.}
Since $\eta_0\in\mathcal B$, we have
$|\eta_0(r)|\le C\omega(r)$. Together with
\eqref{eq:F-derivative-bounds}, this gives, for $1\le|\alpha|\le2$,
\begin{align}
\left|
\partial_q^\alpha F_0(q,r)\eta_0(r)\mu_0(r)
\right|
&\le
C\omega(r)^2\mu_0(r).
\end{align}
The right-hand side is integrable by \eqref{eq:singular-basic} with
$\xi=0$, so differentiation under the integral is justified by dominated convergence, and
\begin{align}
\sup_q
|D_q^2(\mathscr L_0\eta_0)(q)|
&\le
C.
\label{eq:L0-second-derivative}
\end{align}
The function $\mathscr L_0\eta_0$ is radial and vanishes at the origin,
since $F_0(0,r)=0$. Hence its gradient also vanishes at the origin. Taylor's theorem gives the bound below for $|q|\le1$, while for $|q|\ge1$ it follows from the uniform $\mathcal B$-bound. Thus
\begin{align}
|(\mathscr L_0\eta_0)(q)|
&\le
C\omega(q)^2.
\end{align}
Since $|\psi_0(q)|\le C\omega(q)^2$, equation
\eqref{eq:eta-equation} at $\xi=0$ and $m\ge3/4$ imply
\begin{align}
|\eta_0(q)|
&\le
C\omega(q)^2,
\end{align}
which proves the first estimate in \eqref{eq:eta-bounds}.\medskip

\noindent \emph{Comparison with $\xi\ne0$.}
Using the quadratic bound just obtained, we write
\begin{align}
((\mathscr L_\xi-\mathscr L_0)\eta_0)(q)
&=
\int_{\R^3}
\bigl(F_\xi(q,r)-F_0(q,r)\bigr)
\eta_0(r)\mu_\xi(r)\,\dd r
\notag\\
&\quad+
\int_{\R^3}
F_0(q,r)\eta_0(r)
\bigl(\mu_\xi(r)-\mu_0(r)\bigr)\,\dd r.
\label{eq:L-difference}
\end{align}
By \eqref{eq:Fbasic}, the first term is bounded by
\begin{align}
\left| 
\int_{\R^3}
\bigl(F_\xi(q,r)-F_0(q,r)\bigr)
\eta_0(r)\mu_\xi(r)\,\dd r \right| \le C|\xi|\omega(q)
\int_{\R^3}\omega(r)^3\mu_\xi(r)\,\dd r
&\le
C|\xi|\omega(q),
\end{align}
where we used \eqref{eq:singular-basic}. For the second term, we invoke
$|F_0(q,r)\eta_0(r)|\le C\omega(q)\omega(r)^3$, so
\eqref{eq:measure-comparison} gives the same bound. Hence
\begin{align}
\norm{(\mathscr L_\xi-\mathscr L_0)\eta_0}_{\mathcal B}
&\le
C|\xi|.
\label{eq:L-difference-bound}
\end{align}

Subtracting \eqref{eq:eta-equation} at $\xi$ and at $0$ gives
\begin{align}
(m+g^2\mathscr L_\xi)(\eta_\xi-\eta_0)
&=
\psi_\xi-\psi_0
-
g^2(\mathscr L_\xi-\mathscr L_0)\eta_0.
\end{align}
The inverse of $m+g^2\mathscr L_\xi$ is uniformly bounded on
$\mathcal B$ for small $g$. Together with \eqref{eq:psi-pointwise} and
\eqref{eq:L-difference-bound}, this yields
\begin{align}
|\eta_\xi(q)-\eta_0(q)|
&\le
C|\xi|\omega(q),
\end{align}
which proves the second estimate in \eqref{eq:eta-bounds}. In
particular,
\begin{align}
|\eta_\xi(q)|
&\le
C\omega(q)\bigl(\omega(q)+|\xi|\bigr).
\label{eq:eta-pointwise}
\end{align}

\noindent \emph{Identification and scalar estimates.}
The function on the right-hand side of \eqref{eq:x-eta} belongs to
$L^2(\R^3)$ by \eqref{eq:eta-pointwise} and
\eqref{eq:singular-basic}. Multiplying \eqref{eq:eta-equation} by
$g/(|q|\,|q+\xi|)$ gives
\begin{align}
(1+A_\xi)x_\xi
&=
b_\xi,
\qquad
x_\xi(q)
=
\frac{g\eta_\xi(q)}{|q|\,|q+\xi|}.
\end{align}
The invertibility of $1+A_\xi$ from \eqref{eq:A-technical} therefore
proves \eqref{eq:x-eta}.

Using the definitions of $N_\xi$ and $U_\xi$, and the fact that $(1+A_\xi)^{-1}$ is self-adjoint, we obtain
\begin{align}
U_\xi
&=
g^2
\int_{\R^3}
\eta_\xi(q)\mu_\xi(q)\,\dd q,
&
N_\xi
&=
g^2
\int_{\R^3}
\psi_\xi(q)\eta_\xi(q)\mu_\xi(q)\,\dd q.
\label{eq:scalar-integrals}
\end{align}
By \eqref{eq:eta-pointwise} and \eqref{eq:singular-basic}, the first
integral is uniformly bounded, and hence
\begin{align}
|U_\xi|
&\le
Cg^2.
\end{align}

For $N_\xi$, the estimates \eqref{eq:psi-pointwise:1},
\eqref{eq:psi-pointwise}, \eqref{eq:eta-bounds}, and
\eqref{eq:eta-pointwise} give
\begin{align}
|\psi_\xi(q)\eta_\xi(q)-\psi_0(q)\eta_0(q)|
&\le
C|\xi|\omega(q)^2,
&
|\psi_0(q)\eta_0(q)|
&\le
C\omega(q)^4
\le
C\omega(q)^3.
\end{align}
Consequently,
\begin{align}
|N_\xi-N_0|
&\le
g^2
\int_{\R^3}
|\psi_\xi\eta_\xi-\psi_0\eta_0|
\mu_\xi\,\dd q +
g^2
\int_{\R^3}
|\psi_0\eta_0|
|\mu_\xi-\mu_0|\,\dd q
\le 
Cg^2|\xi|,
\end{align}
by \Cref{lem:singular-integrals}.

Finally, \eqref{eq:A-technical} and \eqref{eq:Riesz} imply
\begin{align}
0\le T_\xi
&\le
2\norm{t_\xi}_2^2
=
\frac{2\pi^3g^2}{|\xi|}.
\end{align}
This proves \eqref{eq:scalar-bounds}.
\end{proof}

We can now conclude the proof of \Cref{prop:analytic-estimates}.

\begin{proof}[Proof of \Cref{prop:analytic-estimates}]
The Hilbert--Schmidt and $L^2$ statements follow from
\eqref{eq:HS-and-b}, while the bounds on $A_\xi$ follow from
\eqref{eq:A-technical}. The lower bound on $c_\xi$ and the estimate on
$c_\xi-c_0$ are provided by \Cref{lem:cxi}, and the remaining three
scalar estimates are precisely \eqref{eq:scalar-bounds}.
\end{proof}

\appendix

\section{The Lieb--Yamazaki bound}\label{app:weighted}

The quadratic form estimate for the field operator used throughout the paper is 
\begin{equation}\label{eq:form-bound-full}
 \pm\Phiop(v_g)
 \le\theta\bigl((\Pf-p)^2+N\bigr)
   +C\left(\frac{g^2}{\theta}+\frac{g^4}{\theta^3}\right)
\end{equation}
for every $g\in\R$, $p\in\R^3$, and $0<\theta\le1$.
In particular, the remainder is $Cg^2/\theta$ when $|g|\le\theta$, as stated in \eqref{eq:form-bound}. The estimate goes back to \cite{LiebYamazaki1958}.

For completeness, we provide its derivation.  For $\Lambda\ge1$,
split $v_g=v_{\le\Lambda}+v_{>\Lambda}$, where the Fourier modes are restricted to $\{|k| \le \Lambda\}$ and $\{|k| >  \Lambda\}$, respectively. Since
$\norm{v_{\le\Lambda}}_2^2=4\pi g^2\Lambda$, the annihilation-operator
bound and Young's inequality give, for $\Psi\in\mathcal Q$,
\begin{align}
 |\langle\Psi,\Phiop(v_{\le\Lambda})\Psi\rangle|
 \le\frac\theta2\norm{N^{1/2}\Psi}^2
    +C\frac{g^2\Lambda}{\theta}\norm{\Psi}^2.
\end{align}
For the high-momentum part put
$w_j(k)=k_j|k|^{-2}v_g(k)\one_{\{|k|>\Lambda\}}$.
Then $\sum_j\norm{w_j}_2^2=4\pi g^2/\Lambda$, and 
\begin{align}
 \Phiop(v_{>\Lambda})
 =\sum_{j=1}^3[(\Pf)_j-p_j,a^*(w_j)-a(w_j)].
\end{align}
The usual creation and annihilation estimates therefore imply
\begin{align}
 |\langle\Psi,\Phiop(v_{>\Lambda})\Psi\rangle|
 &\le C\frac{|g|}{\sqrt\Lambda}
   \norm{(\Pf-p)\Psi}\norm{(N+1)^{1/2}\Psi}\notag \\
 &\le\frac\theta2\norm{(\Pf-p)\Psi}^2
    +C\frac{g^2}{\theta\Lambda}
       \bigl(\norm{N^{1/2}\Psi}^2+\norm{\Psi}^2\bigr).
\end{align}
Choose $\Lambda=C_0\max\{1,g^2/\theta^2\}$ with $C_0$ sufficiently large.
The coefficient of $N$ in the last line is then at most
$\theta/2$. Combining the two estimates proves
\eqref{eq:form-bound-full}.\bigskip

\noindent\textbf{Use of AI.} ChatGPT was used during the preparation of this manuscript to assist with parts of the mathematical argument and to provide proofreading and suggestions for improving the exposition. All mathematical arguments and results were independently reviewed and verified by the author, who takes full responsibility for the content of the paper.\medskip

\noindent\textbf{Acknowledgments.} The author thanks Robert Seiringer for helpful discussions, and acknowledges support from the German Research Foundation (DFG) within the scientific network A(E)MP (Grant No.\ 569490025).

\medskip

\noindent\textbf{Data availability.}
No data was generated or analyzed in this work.

\medskip

\noindent\textbf{Conflict of interest.}
The author declares no competing interests.


\begin{thebibliography}{99}

\bibitem{Dahlbaek2017}
J.~Dahlb\ae k,
\emph{Spectral analysis of large particle systems},
PhD thesis, Aarhus University, 2017, Chapter~1 and Theorem~1.3.

\bibitem{GriesemerWuensch2016}
M.~Griesemer and A.~W\"unsch,
\emph{Self-adjointness and domain of the Fr\"ohlich Hamiltonian},
J. Math. Phys. \textbf{57} (2016), 021902.

\bibitem{LampartMitrouskasMysliwy2023}
J.~Lampart, D.~Mitrouskas, and K.~My\'sliwy,
\emph{On the global minimum of the energy--momentum relation for the polaron},
Math. Phys. Anal. Geom. \textbf{26} (2023), Article~17.

\bibitem{LiebYamazaki1958}
E.~H.~Lieb and K.~Yamazaki,
\emph{Ground-state energy and effective mass of the polaron},
Phys. Rev. \textbf{111} (1958), 728--733.

\bibitem{MitrouskasSeiringer2023}
D.~Mitrouskas and R.~Seiringer,
\emph{Ubiquity of bound states for the strongly coupled polaron},
Pure Appl. Anal. \textbf{5} (2023), 973--1008.

\bibitem{Miyao2010}
T.~Miyao,
\emph{Nondegeneracy of ground states in nonrelativistic quantum field theory},
J. Operator Theory \textbf{64} (2010), 207--241.

\bibitem{Moller2006}
J.~S.~M\o ller,
\emph{The polaron revisited},
Rev. Math. Phys. \textbf{18} (2006), 485--517.

\bibitem{Polzer2023}
S.~Polzer,
\emph{Renewal approach for the energy--momentum relation of the Fr\"ohlich polaron},
Lett. Math. Phys. \textbf{113} (2023), Article~90.

\bibitem{Schmidt2026}
T.~Schmidt,
\emph{The Fr\"ohlich Polaron at large total momentum in $d=3$}, arXiv preprint (2026),
\href{https://arxiv.org/abs/2610.16081v1}{arXiv:2610.16081}.

\bibitem{Seiringer2023}
R.~Seiringer,
\emph{Absence of excited eigenvalues for Fr\"ohlich type polaron models at weak coupling},
J. Spectr. Theory \textbf{13} (2023), 1045--1055.

\end{thebibliography}
\end{document}